\documentclass[nohyper,12pt,letterpaper]{JHEP3}

\usepackage{amsmath}

\def\beq{\begin{equation}}
\def\eeq{\end{equation}}

\def\text#1{\mbox{\scriptsize #1}}

\title{Renormalization of Hamiltonian Field Theory; a non-perturbative and
non-unitarity approach}

\author{
{Amir H. Rezaeian$^a\ $ and Niels R. Walet$^b\ $}

\vspace*{5mm}

{\it Department of Physics, UMIST, PO Box 88, Manchester, M60 1QD, UK} \\
$^a\ $ E-mail: \email{Rezaeian@Theory.phy.umist.ac.uk}\\
$^b\ $ E-mail: \email{Niels.Walet@umist.ac.uk}

} 

\abstract{
Renormalization of Hamiltonian field theory is usually a rather painful
algebraic or numerical exercise. By combining a method based on the
many-body coupled cluster method, analyzed in detail by Suzuki and Okamoto,
with a Wilsonian approach to renormalization, we show that a powerful
and elegant method exist to solve such problems. The method is in
principle non-perturbative, and is not necessarily unitary.}

\begin{document}
\section{Introduction}

The power of Hamiltonian methods is well known from the study of non-relativistic
many-body systems and from strongly-interacting few particle systems,
even though a Lagrangian approach is usually chosen for relativistic
theories. Hamiltonian methods for strongly-interacting systems are
intrinsically non-perturbative and usually contain a Tamm-Dancoff
type approximation, in the sense that, in practice, one has to limit
the bound state as an expansion over states containing a small number
of particles. This truncation of the Fock space gives rise to a new
class of non-perturbative divergences, since the truncation does not
allow us to take into account all diagrams for any given order in
perturbation theory. Therefore renormalization issues have to be considered
carefully. Two very different remedies for this issue are the use
of light-front Tamm-Dancoff field theory (LFFT) \cite{3} and the
application of the coupled cluster method (CCM) \cite{4,5}. In the
LFFT the quantization plane is chosen to coincide with the light front,
therefore the divergences that plagued the original theories seem
to disappear \cite{6}. Furthermore, not having to include interactions
in boost operators allows a renormalizable truncation scheme \cite{7}.
One of the most important difficulties in LFFT is the complicated
structure of the renormalization process \cite{8}. In the standard
form of CCM, on the other hand, the amplitudes obey a system of coupled
non-linear equations which contain some ill-defined terms because
of ultraviolet divergences. It has been shown \cite{9} that the ill-defined
amplitudes, which are also called critical topologies, can be systematically
removed, by exploiting the linked-cluster property of the ground state.
This can be done by introducing a mapping which transfers them into
a finite representations without making any approximation such as
a coupling expansion. Thus far this resummation method has been restricted
to superrenormalizable theories due to its complexity.

A natural question in the renormalization within the Hamiltonian formalism
arises, since one could also perform renormalization in a Lagrangian
framework and finally construct the corresponding Hamiltonian by means
of standard Legendre transformation.~{[}It should be noticed that
this is not generally applicable if the Lagrangian contains higher-order
time-derivatives.{]} The Hamiltonian formalism, despite a certain
lack of elegance, has the advantage that it is very economical, and
one can use all the know-how of quantum many-body theory.

In the last decade extensive attempts have been made to give a prescription
for renormalization within the Hamiltonian formalism \cite{aps1,aps2,10}.
Commonly unitary transformations are used to decouple the high- and
low-energy modes aiming at the partial diagonalization of the Hamiltonian.
One of the elegant approach in this context has been the so-called
similarity renormalization group (SRG) proposed by Glazek and Wilson
\cite{aps1} (and by Wegner \cite{aps2} independently). The SRG resembles
the original Wilsonian renormalization group formalism \cite{aps3},
since a transformation that explicitly runs the cutoff is developed.
In an early paper, Wilson \cite{aps3} exploited a similar transformation
which was originally introduced by Bloch \cite{aps4}. There the
Hamiltonian matrix is reduced by using a transformation which lowers
the cutoff initially imposed on the individual states. Later, Wilson
abandoned this formalism in favour of a Lagrangian one. The most important
reason was that the Bloch transformation is ill-defined and produces
fake divergences. These divergences emerge from denominators which
contain a small energy difference between states retained and states
removed by the transformation. The SRG \cite{aps1,aps2} is designed
to be free of such small energy denominators and to eliminate interactions
which change the energies of the unperturbed states by a large amount.
However, there are several issues in this approach: it is hard to
incorporate loop expansions within the method, the SRG can not systematically
remove interactions which change the number of particles (i.e, when
the Hamiltonian is not diagonal in particle number space), and most
importantly, the computations are complex and there is no an efficient
non-perturbative calculable scheme. 

In this paper we introduce a method for obtaining the low energy
effective operators in the framework of a CCM approach. The
transformation constructed avoids the small denominators that plague
old-fashioned perturbation theory. Neither perturbation theory nor
unitarity of the transformation are essential for this method. The method is
non-perturbative, since there is no expansion in the coupling
constant; nonetheless, the CCM can be conceived as a topological
expansion in the number of correlated excitations. We show that
introducing a double similarity transformation using linked-cluster
amplitudes will simplify the partial diagonalization underlying
renormalization in Hamiltonian approaches. However, a price must be
paid: due to the truncation the similarity transformations are not
unitary, and accordingly the hermiticity of the resultant effective
Hamiltonian is not manifest. This is related to the fact that we have
a biorthogonal representation of the given many-body problem. There is
a long tradition of such approaches. The first we are aware of are
Dyson-type bosonization schemes \cite{11}. {[}Here one chooses to map
the generators of a Lie algebra, such that the raising generators have
a particularly simple representation.{]} The space of states is mapped
onto a larger space where the physically realizable states are
obtained by constrained dynamics. This is closely related to CCM
formalism, where the extended phase space is a complex manifold, the
physical subspace constraint function has been shown to be
of second class and the physical shell itself was found to be a
K\"{a}hler manifold \cite{12}. The second is the Suzuki-Lee method in
the nuclear many-body (NMT) problem  \cite{13,Ni}, which reduces the
full many-body problem to a problem in a small configuration space and
introduces a related effective interaction. The effective interaction
is naturally understood as the result of certain transformations which
decouple a model space from its complement. As is well know in the
theory of effective interactions, unitarity of the transformation used
for decoupling or diagonalization is not necessary. Actually, the
advantage of a non-unitarity approach is that it can give a very
simple description for both diagonalization and the ground state.
This has been discussed by many authors \cite{14} and, although it
might lead to a non-hermitian effective Hamiltonian, it has been shown
that hermiticity can be recovered \cite{12,15}. Nevertheless
non-hermiticity is negligible if the model space and its complement
are not strongly correlated \cite{16,17}. Therefore defining a good
model space can in principle control the accuracy of CCM.

To solve the relativistic bound state problem one needs to
systematically and simultaneously decouple 1) the high-energy from
low-energy modes and 2) the many- from the few-particle states. We
emphasize in this paper that CCM can in principle be an adequate
method to attack both these requirements. Our hope is to fully utilize
Wilsonian Exact renormalization group \cite{18} within the CCM
formalism. Here the high energy modes will be integrated out leading
to a modified low-energy Hamiltonian in an effective many-body space.
Notice that our formulation does not depend on the form of dynamics
and can be used for any quantization scheme, e.g., equal time or
light-cone.

The organization of this paper is as follows. In section II, we discuss
our approach and it's foundation. Finally we conclude and present an outlook
in section III.

\section{Formalism}

The discussion in this section is partially based on the work of Suzuki
and Okamoto \cite{Ni}. Let us consider a system described by a Hamiltonian
$H(\Lambda)$ which has, at the very beginning, a large cut-off $\Lambda$.
We assume that the renormalized Hamiltonian $H^{\text{eff}}(\Lambda)$
up to scale $\Lambda$ can be written as the sum of the canonical
Hamiltonian and a {}``counterterm'' $H_{C}(\Lambda)$, \begin{equation}
H^{\text{eff}}(\Lambda)=H(\Lambda)+H_{C}(\Lambda).\end{equation}
 Our aim is to construct the renormalized Hamiltonian by obtaining
this counterterm. Now imagine that we restrict the Hamiltonian to
a lower energy scale $(\mu)$, where we want to find an effective
Hamiltonian $H^{\text{eff}}(\mu)$ which has the same energy spectrum
as the original Hamiltonian in the smaller space. Formally, we wish
to transform the Hamiltonian to a new basis, where the medium-energy
modes $\mu<k<\Lambda$, decouple from the low-energy ones, while the
low-energy spectrum remains unchanged. We define two subspaces, the
intermediate-energy space $Q$ containing modes with $\mu<k<\Lambda$
and a low-energy space $P$ with $\mu\leq k$. Our renormalization
approach is based on decoupling of the complement space $Q$ from
the model space $P$. Thereby the decoupling transformation generates
a new effective interaction $\delta H(\mu,\Lambda)$ containing the
physics between the scale $\Lambda$ and $\mu$. One can then determine
the counterterm by requiring coupling coherence \cite{10,coh}, namely
that the transformed Hamiltonian has the same form as the original
one but with $\Lambda$ every where replaced by $\mu$. {[}This is
in contrast to the popular Effective Field Theory approach, where one includes
all permissible couplings of a given order and fixes them by requiring
observable computed be both cutoff-independent and Lorentz covariant.{]} The operators
$P$ and $Q$ which project a state onto the model space and its complement,
satisfy $P^{2}=P$, $Q^{2}=Q$, $PQ=0$ and $P+Q=1$. We introduce
an isometry operator $G$ which maps states in the $P$- onto the
$Q$- space, 
\begin{equation}
|q\rangle=G|p\rangle\hspace{1cm}(|q\rangle\in Q,|p\rangle\in P).
\end{equation}
 The operator $G$ is the basic ingredient in a family of {}``integrating-out
operators'', which passes information about the correlations of the
high energy modes to the low-energy space. The operator $G$ obeys
$G=QGP$, $GQ=0$, $PG=0$ and $G^{n}=0$ for $n\geqslant2$. The
rather surprising direction for $G$ to act in is due to the definition
Eq.~(\ref{eq4}) below (cf. the relation between the active and passive
view of rotations). In order to give a general form of the effective
low-energy Hamiltonian, we define another operator $X(n,\mu,\Lambda)$,
\begin{equation}
X(n,\mu,\Lambda)=(1+G)(1+G^{\dag}G+GG^{\dag})^{n}.\label{eq2}\end{equation}
 ($n$ is a real number.) The inverse of $X(n,\mu,\Lambda)$ can be
obtained explicitly, \begin{equation}
X^{-1}(n,\mu,\Lambda)=(1+G^{\dag}G+GG^{\dag})^{-n}(1-G).\label{eq3}\end{equation}
 The special case $n=0$ is equivalent to the transformation introduced
in Ref.~\cite{25} to relate the hermitian and non-hermitian effective
operators in the energy-independent Suzuki-Lee approach. We now consider
the transformation of $H(\Lambda)$ defined as 
\begin{equation}
\overline{H}(n,\mu,\Lambda)=X^{-1}(n,\mu,\Lambda)H(\Lambda)X(n,\mu,\Lambda),\label{eq4}\end{equation}
 where we have \begin{equation}
H(\Lambda)\rightarrow\overline{H}(n,\mu,\Lambda)\equiv H(\mu)+\delta H(\mu,\Lambda).\end{equation}
 One can prove that if $\overline{H}(n,\mu,\Lambda)$ satisfies the
desirable decoupling property, \begin{equation}
Q\overline{H}(n,\mu,\Lambda)P=0,\label{eq5}\end{equation}
 or more explicitly, by substituting the definition of $X(n,\mu,\Lambda)$
and $X^{-1}(n,\mu,\Lambda)$ from Eqs.~(\ref{eq2}-\ref{eq3}), \begin{equation}
QH(\Lambda)P+QH(\Lambda)QG-GPH(\Lambda)P-GPH(\Lambda)QG=0,\label{eq6}\end{equation}
 that $H^{\text{eff}}(\mu)\equiv{\mathcal{H}}(n,\mu)=P\overline{H}(n,\mu,\Lambda)P$
is an effective Hamiltonian for the low energy degrees of freedom.
In other words, it should have the same low-energy eigenvalues as
the original Hamiltonian. The proof is as follows:

Consider an eigenvalue equation in the $P$ space with $\{|\phi(k)\rangle\in P\}$,
\begin{eqnarray}
 &  & P\overline{H}(n,\mu,\Lambda)P|\phi(k)\rangle=E_{k}PX^{-1}(n,\mu,\Lambda)X(n,\mu,\Lambda)P|\phi(k)\rangle.\label{eq7}\end{eqnarray}
 By multiplying both sides by $X(n,\mu,\Lambda)$ and making use of
the decoupling property Eq.~(\ref{eq5}), we obtain \begin{equation}
H(\Lambda)X(n,\mu,\Lambda)P|\phi(k)\rangle=E_{k}X(n,\mu,\Lambda)P|\phi(k)\rangle.\label{eq8}\end{equation}
 This equation means that $E_{k}$ in Eq.~(\ref{eq7}) agrees with
one of the eigenvalue of $H(\Lambda)$ and $X(n,\mu,\Lambda)P|\phi(k)\rangle$
is the corresponding eigenstate. Now we demand that \begin{equation}
H^{\text{eff}}(\mu)\equiv H(\mu)+H_{C}(\mu).\end{equation}
 This requirement uniquely determines the counterterm $H_{C}$.

In the same way we can also obtain the $Q$-space effective Hamiltonian,
from the definition of $\overline{H}(n,\mu,\Lambda)$. It can be seen
that if $G$ satisfies the requirement in Eq.~(\ref{eq6}), then
we have additional decoupling condition \begin{equation}
P\overline{H}(n,\mu,\Lambda)Q=0.\label{decoupling}\end{equation}
 We will argue later that \emph{both} of the decoupling conditions
Eqs.~(\ref{eq5}) and (\ref{decoupling}) are necessary in order to
have a sector-independent renormalization scheme. The word {}``sector''
here means the given truncated Fock space. Let us now clarify the
meaning of this concept. To maintain the generality of the previous
discussion, we use here the well known Bloch-Feshbach formalism \cite{aps4,b,fesh}.
The Bloch-Feshbach method exploits projection operators in the Hilbert
space in order to determine effective operators in some restricted
model space. This technique seems to be more universal than Wilson's
renormalization formulated in a Lagrangian framework. This is due
to the fact that in the Bloch-Feshbach formalism, other irrelevant
degrees of freedom (such as high angular momentum, spin degrees of
freedom, number of particles, etc.) can be systematically eliminated
in the same fashion.

Assume that the full space Schr\"{o}dinger equation is $H(\Lambda)|\psi\rangle=E|\psi\rangle$
and for simplicity $|\psi\rangle$ has been normalized to one. We
explicitly construct the energy dependent effective Hamiltonian in this formalism,
\begin{eqnarray}
H^{\text{eff}} & = & P\overline{H}P+P\overline{H}Q\frac{1}{E-Q\overline{H}Q}Q\overline{H}P,\label{F1}\end{eqnarray}
 where $\overline{H}$ can be a similarity transformed Hamiltonian.
This equation resembles the one for Brueckner's reaction matrix (or {}``G''-matrix)
equation in nuclear many-body theory (NMT). In the same way for arbitrary
operator $O$ (after a potential similarity transformation), we construct
the effective operator \begin{eqnarray}
O^{\text{eff}} & = & P\overline{O}P+P\overline{H}Q\frac{1}{E-Q\overline{H}Q}Q\overline{O}P+P\overline{O}Q\frac{1}{E-Q\overline{H}Q}Q\overline{H}P\nonumber \\
 & + & P\overline{H}Q\frac{1}{E-Q\overline{H}Q}Q\overline{O}Q\frac{1}{E-Q\overline{H}Q}Q\overline{H}P.\label{F2}\end{eqnarray}
 The $E$-dependence in Eqs.~(\ref{F1}) and (\ref{F2}) emerges
from the fact that the effective interaction in the reduced space
is not assumed to be decoupled from the excluded space. However, by
using the decoupling conditions introduced in Eqs.~(\ref{eq5}) and
(\ref{decoupling}), we observe that energy dependence can be removed,
and the effective operators become \begin{eqnarray}
H^{\text{eff}} & = & P\overline{H}P={\mathcal{H}}(n,\mu),\nonumber \\
O^{\text{eff}} & = & P\overline{O}P={\mathcal{O}}(n,\mu).\label{final}\
\end{eqnarray}
 The decoupling property makes the operators in one sector independent
of the other sector. The effects of the excluded sector is taken into
account by imposing the decoupling conditions. This is closely related
to the folded diagram method in NMT for removing energy-dependence
\cite{folded}. (It is well-known in NMT that $E$-dependence in the
$G$-matrix emerges from non-folded diagrams which can be systematically
eliminated using the effective interaction approach). The above argument
was given without assuming an explicit form for $X$ and thus the
decoupling conditions are more fundamental than the prescription used
to derive these conditions. We will show later that one can choose
a transformation $X$, together with the model space and its complement,
which avoids the occurrence of {}``small energy denominators''.
We now show that Lorentz covariance in a given sector does not hinge
on a special form of similarity transformation. We assume ten Poincar\'{e}
generators $L_{i}$ satisfying \begin{equation}
[L_{i},L_{j}]=\sum a_{ij}^{k}L_{k},\end{equation}
 where the $a_{ij}^{k}$ are the known structure coefficients. One
can show that if the operators $L_{i}$ satisfy the decoupling conditions~
$Q\bar{L}_{i}P=0$ and $P\bar{L}_{i}Q=0$ then it follows that \begin{equation}
[L_{i}^{\text{eff}},L_{j}^{\text{eff}}]=\sum a_{ij}^{k}L_{k}^{\text{eff}}.\end{equation}
 This leads to a relativistic description even after simultaneously
integrating out the high-frequency modes and reducing the number of
particles. Indeed we conjecture that requiring the decoupling conditions
makes the effective Hamiltonian free of Lorentz-noninvariant operators
for a given truncated sector regardless of the regularization scheme.
However, one may still be faced with an effective Hamiltonian which
violates gauge invariance (for e.g., when sharp cutoff is employed).

Note that the solution to Eq.~(\ref{eq6}) is independent of the number
$n$. One can make use of Eq.~(\ref{eq6}) and its complex conjugate to
show that for any real number $n$, the following relation for the
effective low-energy Hamiltonian \begin{equation}
{\mathcal{H}}(n,\mu)=\mathcal{H}^{\dag}(-n-1,\mu).\label{eq10}\end{equation}
The case $n=-1/2$ is special since the effective Hamiltonian is
hermitian, \begin{equation}
{\mathcal{H}}(-1/2,\mu)=(P+G^{\dag}G)^{1/2}H(\Lambda)(P+G)(P+G^{\dag}G)^{-1/2}.\label{hermi}\end{equation}
Hermiticity can be verified from the relation
\cite{26}
\begin{equation}
e^{T}P=(1+G)(P+G^{\dag}G)^{-1/2},\end{equation} where,
\begin{equation}
T=\arctan(G-G^{\dag})=\sum_{n=0}^{\infty}\frac{(-1)^{n}}{2n+1}\big(G(G^{\dag}G)^{n}-\textrm{h.c}.).
\end{equation}
Since the operator $T$ is anti-hermitian, $e^{T}$ is a unitary
operator. From the above expression Eq.~(\ref{hermi}) can be written
in the explicitly hermitian form \begin{equation}
{\mathcal{H}}(-1/2,\mu)=Pe^{-T}H(\Lambda)e^{T}P.\end{equation} As was
already emphasized, renormalization based on unitary transformations
is more complicated and non-economical. Thus we will explore a
non-unitary approach. An interesting non-hermitian effective
low-energy Hamiltonian can be obtained for $n=0$, \begin{equation}
{\mathcal{H}}(0,\mu,\Lambda)=PH(\Lambda)(P+QG).\label{eq11}\end{equation}
This form resembles the Bloch and Horowitz type of effective
Hamiltonian as used in NMT \cite{b}, and was used by Wilson in quantum
field theory \cite{aps4}. In the context of the CCM, this form
leads to the folded diagram expansion well known in many-body theory
\cite{f}. It is of interest that various effective low-energy Hamiltonians
can be constructed according to Eq.~(\ref{final}) by the use of
the mapping operator $G$ which all obey the decoupling property Eq.~(\ref{eq5})
and Eq.~(\ref{decoupling}). Neither perturbation theory nor hermiticity
is essential for this large class of effective Hamiltonians.

The CCM approach is, of course, just one of the ways of describing the
relevant spectrum by means of non-unitary transformations. According
to our prescription the model space is
$P:\{|L\rangle\bigotimes|0,b\rangle_{h},L\leq\mu\}$, where
$|0,b\rangle_{h}$ is a bare high energy vacuum (the ground state of
high-momentum of free-Hamiltonian) which is annihilated by all the
high frequency annihilation operators $\{ C_{I}\}$ (for a given
quantization scheme, e.g., equal time or light-cone) , the set of
indices $\{ I\}$ therefore defines a subsystem, or cluster, within the
full system of a given configuration. The actual choice depends upon
the particular system under consideration. In general, the operators
$\{ C_{I}\}$ will be products (or sums of products) of suitable
single-particle operators. We assume that the annihilation and its
corresponding creation $\{ C_{I}^{\dag}\}$ subalgebras and the state
$|0,b\rangle_{h}$ are cyclic, so that the linear combination of state
$\{ C_{I}^{\dag}|0,b\rangle_{h}\}$ and $\{~_{h}\langle b,0|C_{I}\}$
span the decoupled Hilbert space of the high-momentum modes,
$\{|H\rangle\}$, where $\mu<H<\Lambda$.  It is also convenient, but
not necessary, to impose the orthogonality condition,
$\langle0|C_{I}C_{J}^{\dag}|0\rangle=\delta(I,J)$, where $\delta(I,J)$
is a Kronecker delta symbol. The complement space is
$Q:\{|L\rangle\bigotimes\left(|H\rangle-|0,b\rangle_{h}\right)\}$.
Our main goal is to decouple the $P$-space from the $Q$-space. This
gives sense to the partial diagonalization of the high-energy part of
the Hamiltonian. The states in full Hilbert space are constructed by
adding correlated clusters of high-energy modes onto the $P$-space, or
equivalently integrating out the high-energy modes from the
Hamiltonian,
\begin{eqnarray}
 &  & |f\rangle=X(\mu,\Lambda)|p\rangle=e^{\hat{S}}e^{-\hat{S}'}|0,b\rangle_{h}\bigotimes|L\rangle=e^{\hat{S}}|0,b\rangle_{h}\bigotimes|L\rangle,\label{eq13}\\
 &  & \langle\widetilde{f}|=\langle L|\bigotimes~_{h}\langle b,0|X^{-1}(\mu,\Lambda)=\langle L|\bigotimes~_{h}\langle0|e^{\hat{S}'}e^{-\hat{S}},\label{eq14}\end{eqnarray}
 where the operators $X(\mu,\Lambda)$ and $X^{-1}(\mu,\Lambda)$
have been expanded in terms of independent coupled cluster excitations
$I$, \begin{eqnarray}
 &  & \hat{S}=\sum_{m=0}\hat{S}_{m}\left(\frac{\mu}{\Lambda}\right)^{m},\hspace{2cm}\hat{S}_{m}=\sideset{}{'}\sum_{I}\hat{s}_{I}C_{I}^{\dag},\nonumber \\
 &  & \hat{S}'=\sum_{m=0}\hat{S}'_{m}\left(\frac{\mu}{\Lambda}\right)^{m},\hspace{2cm}\hat{S}'_{m}=\sideset{}{'}\sum_{I}{}\hat{s}'_{I}C_{I}.
\end{eqnarray}
 Here the primed sum means that at least one fast particle is created
 or destroyed $(I\neq0)$, and momentum conservation in $P\bigoplus Q$
 is included in $\hat{s}_{I}$ and
 $\hat{s}'_{I}$. $\hat{S}_{m}(\hat{S}'_{m})$ are not generally
 commutable in the low-energy Fock space, whereas they are by
 construction commutable in the high-energy Fock space.  It is
 immediately clear that states in the interacting Hilbert space are
 normalized, $\langle\widetilde{f}|f\rangle=~_{h}\langle
 b,0|0,b\rangle_{h}=1$.  We have two types of parameters in this
 procedure: One is the coupling constant of the theory ($\lambda$),
 and the other is the ratio of cutoffs ($\mu/\Lambda$). The explicit
 power counting makes the degree of divergence of each order smaller
 than the previous one. According to our logic, Eq.~(\ref{eq11}) can
 be written as \begin{equation} {\mathcal{H}}(\mu)=~_{h}\langle
 b,0|X^{-1}(\mu,\Lambda)H(\Lambda)X(\mu,\Lambda)|0,b\rangle_{h},\label{eq15}\end{equation}
 with $X(\mu,\Lambda)$ and $X^{-1}(\mu,\Lambda)$ defined in
 Eq.~(\ref{eq13}) and Eq.~(\ref{eq14}). We require that effective
 Hamiltonian ${\mathcal{H}}(\mu)$ obtained in this way remains form
 invariant or coherent \cite{10,coh}.  This requirement satisfies on
 an infinitely long renormalization group trajectory and thus does
 constitutes a renormalized Hamiltonian. Thereby one can readily
 identify the counter terms produced from expansion of
 Eq.~(\ref{eq15}).

It is well-know in many-body applications that the exponential Ansatz
Eqs.~(\ref{eq13}) and (\ref{eq14}) guarantees automatically proper
size-exclusivity and conformity with the Goldstone linked-cluster
theorem to all level of truncation. This parametrization does not
manifestly preserve hermitian conjugacy. However, it is compatible
with the Hellmann-Feynman theorem (HFT), in other words, demanding
hermiticity will violate this theorem at any level of truncation.
On the other hand, with this parametrization the phase space $\{\hat{s}_{I},\hat{s}'_{I}\}$
for a given $m$ is a symplectic differentiable manifold. Thereby
all the geometrical properties of the configuration space can be precisely
defined.~(In a more ordinary language, the canonical equations of
motion with respect to phase space define a set of trajectories,
which fill the whole dynamically allowed region of the phase space.)
There is a deep connection between these three properties and we can
not give up one without losing something else as well \cite{27}.
The individual amplitudes for a given $m$, $\{\hat{s}_{I}^{m},\hat{s}'{}_{I}^{m}\}\equiv\{\hat{s}_{I},\hat{s}'_{I}\}_{m}$,
can generally be functionals of the low- and high-energy field operators
and have to be fixed by the dynamics of quantum system. This is a
complicated set of requirements. However, we require less than that.
Suppose that after a similar transformation of Hamiltonian, $\overline{H}$,
we obtain an effective Hamiltonian of the form \begin{equation}
\overline{H}=H(\text{low})+H_{\text{free}}(\text{high})+C_{I}^{\dag}V_{IJ}C_{J},\label{eqH}\end{equation}
 where $V_{IJ}$ is an arbitrary operator in the low frequency space.
The $I$ and $J$ indices should be chosen such that the last term
in Eq.~(\ref{eqH}) contains at least one creation- operator and
one annihilation-operator of high frequency. By using Rayleigh-Schr\"{o}dinger
perturbation theory, it can be shown that the free high-energy vacuum
state of $H_{\text{free}}(\text{high})$ is annihilated by Eq.~(\ref{eqH})
and remains without correction at any order of perturbation theory.
Having said that, we will now consider how to find the individual
amplitudes $\{\hat{s}_{I},\hat{s}'_{I}\}_{m}$ that transfer the Hamiltonian
into the form Eq.~(\ref{eqH}). We split the Hamiltonian in five
parts: \begin{equation}
H=H_{1}+H_{2}^{\text{free}}(\text{high})+V_{C}(C_{I}^{\dag})+V_{A}(C_{I})+V_{B},\label{eq16}\end{equation}
 where $H_{1}$ contains only the low frequency modes with $k\leq\mu$,
$H_{2}$ is the free Hamiltonian for all modes with $\mu<k<\Lambda$,
$V_{C}$ contains low frequency operators and products of the high
frequency creation operators $C_{I}^{\dag}$ and $V_{A}$ is the hermitian
conjugate of $V_{C}$. The remaining terms are contained in $V_{B}$,
these terms contain at least one annihilation and creation operators
of the high energy modes. Our goal is to eliminate $V_{C}$ and $V_{A}$
since $V_{B}$ annihilates the vacuum. The ket-state coefficients
$\{\hat{s}_{I}\}_{m}$ are worked out via the ket-state Schr\"{o}dinger
equation $H(\Lambda)|f\rangle=E|f\rangle$ written in the form \begin{equation}
\langle0|C_{I}e^{-\hat{S}}He^{\hat{S}}|0\rangle=0,\hspace{2cm}\forall I\neq0.\label{eq17}\end{equation}
 The bra-state coefficients $\{\hat{s}_{I},\hat{s}'_{I}\}_{m}$ are
obtained by making use of the Schr\"{o}dinger equation defined for
the bra-state,~$\langle\widetilde{f}|H(\Lambda)=\langle\widetilde{f}|E$.
First we project both sides on $C_{I}^{\dag}|0\rangle$, then we eliminate
$E$ by making use of the ket-state equation projection with the state
$\langle0|e^{\hat{S}'}C_{I}^{\dag}$ to yield the equations 
\begin{equation}
\langle0|e^{\hat{S}'}e^{-\hat{S}}[H,C_{I}^{\dag}]e^{\hat{S}}e^{-\hat{S}'}|0\rangle=0,\hspace{1cm}\forall I\neq0.\label{eq18}\end{equation}
 Alternatively one can in a unified way apply $e^{\hat{S}}e^{-\hat{S}'}C_{I}^{\dag}|0\rangle$
on the Schr\"{o}dinger equation for the bra-state and obtain \begin{equation}
\langle0|e^{\hat{S}'}e^{-\hat{S}}He^{\hat{S}}e^{-\hat{S}'}C_{I}^{\dag}|0\rangle=0,\hspace{2cm}\forall I\neq0.\label{eq19}\end{equation}
 Equation (\ref{eq17}) and Eqs.~(\ref{eq18}) or (\ref{eq19}) provide
two sets of formally exact, microscopic, operatorial coupled non-linear
equations for the ket and bra. One can solve the coupled equations
in Eq.~(\ref{eq17}) to work out $\{\hat{s}_{I}\}_{m}$ and then
use them as an input in Eqs.~(\ref{eq18}) or (\ref{eq19}).

It is important to notice that Eqs.~(\ref{eq17}) and (\ref{eq18})
can be also derived by requiring that the effective low-energy Hamiltonian
defined in Eq.~(\ref{eq15}), be stationary (i.e. $\delta\mathcal{H}(\mu)=0$)
with respect to all variations in each of the independent functional
$\{\hat{s}_{I},\hat{s}'_{I}\}_{m}$. One can easily verify that the
requirements $\delta\mathcal{H}(\mu)/\delta\hat{s}_{I}=0$ and $\delta\mathcal{H}(\mu)/\delta\hat{s}'_{I}=0$
yield Eqs.~(\ref{eq17}) and (\ref{eq18}). The combination of
Eqs.~(\ref{eq17}) and (\ref{eq18}) does not manifestly satisfy
the decoupling property as set out in Eqs.~(\ref{eq5}) and (\ref{decoupling}).
On the other hand Eqs.~(\ref{eq17}) and (\ref{eq19}) satisfy these
conditions. Equations (\ref{eq17}) and (\ref{eq19}) imply that all
interactions including creation and annihilation of fast particles
({}``$I$'') are eliminated from the transformed Hamiltonian $\mathcal{H(\mu)}$
in Eq.~(\ref{eq15}). In other words, these are decoupling conditions
leading to the elimination of $V_{C}$ and $V_{A}$ from Eq.~(\ref{eq16}),
which is, in essence, a block-diagonalization. Therefore it makes
sense for our purpose to use Eqs.~(\ref{eq17}) and (\ref{eq19})
for obtaining the unknown coefficients, losing some of the elegance
of the CCM elsewhere.

So far everything has been introduced rigorously without invoking
any approximation. In practice one needs to truncate both sets of
coefficients $\{\hat{s}_{I},\hat{s}'_{I}\}_{m}$ at a given order
of $m$. A consistent truncation scheme is the so-called SUB($n,m$)
scheme, where the $n$-body partition of the operator $\{\hat{S},\hat{S}'\}$
is truncated so that one sets the higher partition with $I>n$ to
zero at a given accuracy $m$. Notice that, Eqs.~(\ref{eq18}) and
(\ref{eq19}) provide two equivalent sets of equations in the exact
form, however after the truncation they can in principle be different.
Eqs.~(\ref{eq17}) and (\ref{eq19}) are compatible with the decoupling
property at any level of the truncation, whereas Eqs.~(\ref{eq17})
combined with (\ref{eq18}) are fully consistent with HFT at any level
of truncation. Thus the low-energy effective form of an arbitrary
operator can be computed according to Eq.~(\ref{final}) in the same
truncation scheme used for the renormalization of the Hamiltonian.
In particular, we will show that only in the lowest order ($m=0$),
equations (\ref{eq18}) and (\ref{eq19}) are equivalent, independent
of the physical system and the truncation scheme.

Although our method is non-perturbative, perturbation theory can be
recovered from it. In this way, its simple structure for loop expansion
will be obvious and we will observe that at lower order hermiticity
is preserved. Now we illustrate how this is realizable in our approach.
Assume that $V_{C}$ and $V_{A}$ are of order $\lambda$, we will
diagonalize the Hamiltonian, at leading order in $\lambda$ up to
the desired accuracy in $\mu/\Lambda$. We use the commutator-expansion
\begin{equation}
e^{-S}He^{S}=H+[H,S]+\frac{1}{2!}\big[[H,S],S]+...\quad.\label{eq20}\end{equation}
 Eq.~(\ref{eq17}) can be organized perturbatively in order of $m$,
aiming at elimination of the high momenta degree of freedom up to
the first order in the coupling constant, thus yields 
\begin{eqnarray}
 &  & m=0:\langle0|C_{I}(V_{C}+[H_{2},\hat{S}_{0}])|0\rangle=0,\nonumber \\
 &  & m=1:\langle0|C_{I}([H_{1},\hat{S}_{0}]+[H_{2},\hat{S}_{1}]+[V_{A},\hat{S}_{1}]+[V_{C},\hat{S}_{1}])|0\rangle=0,\nonumber \\
 &  & \hspace{2cm}\vdots\nonumber \\
 &  & m=n:\langle0|C_{I}([H_{1},\hat{S}_{n-1}]+[H_{2},\hat{S}_{n}]+[V_{A},\hat{S}_{n}]+[V_{C},\hat{S}_{n}])|0\rangle=0,\label{eq21}
\end{eqnarray}
 where $I\neq0$. Notice that $\hat{S}_{0}$ is chosen to cancel $V_{C}$
in the effective Hamiltonian, hence it is at least of order of $\lambda$,
consequently it generates a new term $[H_{1},\hat{S}_{0}]$ which is of
higher order in $\mu/\Lambda$ and can be canceled out on the next
orders by $\hat{S}_{1}$. The logic for obtaining the equations above is
based on the fact that $\hat{S}_{n}$ should be smaller than $\hat{S}_{n-1}$
(for sake of convergence) and that the equations should be consistent
with each other. Since $H_{2},V_{A},V_{C}\approx\Lambda$ and $H_{1}\approx\mu$,
from Eq.~(\ref{eq21}) we have the desired relation $\hat{S}_{n}\approx\frac{\mu}{\Lambda}\hat{S}_{n-1}$.
The same procedure can be applied for Eq.~(\ref{eq19}) which leads
to the introduction of a new series of equations in order of $m$,
\begin{eqnarray}
 &  & m=0:\langle0|(V_{A}-[H_{2},\hat{S}'_{0}])C_{I}^{\dag}|0\rangle=0,\nonumber \\
 &  & m=1:\langle0|([H_{1},\hat{S}'_{0}]+[H_{2},\hat{S}'_{1}]+[V_{C},\hat{S}'_{1}]+[V_{A},\hat{S}'_{1}]-[V_{A},\hat{S}_{1}])C_{I}^{\dag}|0\rangle=0,\nonumber \\
 &  & \hspace{2cm}\vdots\nonumber \\
 &  & m=n:\langle0|([H_{1},\hat{S}'_{n-1}]+[H_{2},\hat{S}'_{n}]+[V_{C},\hat{S}'_{n}]+[V_{A},\hat{S}'_{n}]-[V_{A},\hat{S}_{n}])C_{I}^{\dag}|0\rangle=0.\label{eq23}\end{eqnarray}
 Alternatively, we can use Eq.~(\ref{eq18}) to yield the equations
\begin{eqnarray}
 &  & m=0:\langle0|\big([V_{A},C_{I}^{\dag}]-\big[[H_{2},C_{I}^{\dag}],\hat{S}'_{0}\big]\big)|0\rangle=0,\nonumber \\
 &  & m=1:\langle0|\big(\big[[V_{A},C_{I}^{\dag}],\hat{S}_{1}\big]-\big[[H_{2},C_{I}^{\dag}],\hat{S}'_{1}\big]-\big[[V_{A},C_{I}^{\dag}],\hat{S}'_{1}\big]\big)|0\rangle=0,\nonumber \\
 &  & \hspace{2cm}\vdots\nonumber \\
 &  & m=n:\langle0|\big(\big[V_{A},C_{I}^{\dag}],\hat{S}_{n}\big]-\big[[H_{2},C_{I}^{\dag}],\hat{S}'_{n}\big]-\big[[V_{A},C_{I}^{\dag}],\hat{S}'_{n}\big]\big)|0\rangle=0.\label{eq24}\end{eqnarray}
 It is obvious that at order $m=0$, Eqs.~(\ref{eq23}) and (\ref{eq24})
are the same and $\hat{S}'_{0}=\hat{S}_{0}^{\dag}$, which indicates that the
similarity transformation at this level remains unitary. It should
be noted that diagonalization at first order in the coupling constant
introduces a low-energy effective Hamiltonian in Eq.~(\ref{eq15})
which is valid up to the order $\lambda^{3}$. In the same way, diagonalization
at second order in $\lambda$ modifies the Hamiltonian at order $\lambda^{4}$
and leads generally to a non-unitarity transformation. In this way
one can proceed to diagonalize the Hamiltonian at a given order in
$\lambda$ with desired accuracy in $\mu/\Lambda$ . Finally, the
renormalization process is completed by introducing the correct $Z(\Lambda)$
factors which redefine the divergences emerging from Eq.~(\ref{eq15}).

\section{Conclusion and outlook}

In this paper we have outlined a strategy to derive effective
renormalized operators in the Hamiltonian formulation of a field
theory. An example of the application of the method can be found in
Ref.~\cite{plb}. The effective low-frequency operator is obtained by
the condition that it should exhibit decoupling between the low- and
high-frequency degrees of freedom. We showed that the similarity
transformation approach to renormalization can be systematically
classified. The non-hermitian formulation gives a very simple
description of decoupling, leading to a partial diagonalization of the
high-energy part. In the recent paper \cite{plb}, we showed that
non-unitarity representation inherent in our formulation is in favour
of economic computation and does not produce any non-hermiticity in
the relevant terms. The techniques proposed are known from the coupled
cluster many-body theory and invoke neither perturbation nor unitarity
transformation. We showed that our formalism can be solved
perturbatively. In this way, it was revealed that diagonalization at
first order in coupling constant defines a correct low-energy
effective Hamiltonian which is valid up to the order
$\lambda^{3}$. One can show that the non-hermiticity of the effective
Hamiltonian is controllable and might appear in higher order which is
beyond our approximation or in irrelevant terms which can be ignored
in renormalization group sense.

One of the key features which has not yet been exploited is the
non-perturbative aspect of the method; it may well be able to obtain
effective degrees of freedom that are very different from the ones
occurs at the high-energy scale. This is a promising avenue for future
work.

\section{Acknowledgment}
One of the authors (AHR) acknowledges support from British Government
ORS award and UMIST grant. The work of NRW is supported by the UK
engineering and physical sciences research council under grant GR/N15672.

\end{document}